\documentclass[onecollarge,natbib]{svjour2}
\bibpunct{[}{]}{;}{n}{}{,} 
\smartqed  
\usepackage{graphicx}

\usepackage[dvips]{color}
\usepackage{subfigure}

\journalname{Few Body Systems}
\begin{document}

\title{Chiral condensate and Mott-Anderson freeze-out\thanks{Presented at 
the workshop "30 years of strong interactions", Spa, Belgium, 6-8 April 2011.}
}

\titlerunning{Mott-Anderson freeze-out}        

\author{  D.~Blaschke      \and
        J.~Berdermann \and
        J.~Cleymans \and
        K.~Redlich
}


\institute{D.~Blaschke \at
              Institute for Theoretical Physics, University of Wroclaw,
              50-204 Wroclaw, Poland \\
              Bogoliubov  Laboratory of Theoretical Physics, JINR,
              141980  Dubna, Russia \\
              Tel.: +48-71-375 9252\\
              Fax: +48-71-321 4454\\
              \email{blaschke@ift.uni.wroc.pl}           
           \and
           J.~Berdermann \at
              DESY Zeuthen, D-15738 Zeuthen, Germany
           \and
           J.~Cleymans \at
              UCT-CERN Research Centre and Department of Physics,
              Rondebosch 7701, Cape Town, South Africa
           \and
           K.~Redlich \at
              Institute for Theoretical Physics, University of Wroclaw,
              50-204 Wroclaw, Poland \\
              ExtreMe Matter Institute EMMI, GSI, D-64291 Darmstadt, Germany
}

\date{Received: date / Accepted: date}

\maketitle

\begin{abstract}
We present the idea of a Mott-Anderson freeze-out that suggests a key role
of the localization of the hadron wave functions when traversing the
hadronization transition.
The extension of hadron wave functions in dense matter is governed by the
behavior of the chiral quark condensate such that its melting at finite
temperatures and chemical potentials entails an increase of the size of hadrons
and thus their geometrical strong interaction cross sections.
It is demonstrated within a schematic resonance gas model, that a kinetic
freeze-out condition reveals a correlation with the reduction of the chiral
condensate in the phase diagram up to 50\% of its vacuum value.
Generalizing the description of the chiral condensate by taking into account
a full hadron resonance gas such  correlation gets distorted.
We discuss, that this may be due to our approximations in calculating the 
chiral condensate which disregard both, in-medium effects on hadron masses and  
hadron-hadron interactions. The latter, in particular due to quark exchange 
reactions,  could  lead to a delocalization of the hadron wave functions in 
accordance with the picture of a Mott-Anderson transition.

\keywords{Chiral condensate \and chemical freeze-out \and Mott-Anderson
transition}
\end{abstract}

\section{Introduction}
The investigation of the structure of the QCD phase-diagram is one of the goals 
of heavy-ion collision experiments as well as theoretical studies within lattice
QCD (LQCD)  and effective field theory approaches to the non-perturbative
sector of QCD.
Of particular interest are the conditions under which the approximate
chiral symmetry of the QCD Lagrangian  get restored and whether this
transition is to be necessarily accompanied by the deconfinement of quarks and
gluons.
More detailed questions to the QCD phase transitions concern their order, their 
critical exponents as well as the  existence of a critical point or even a 
triple point in the phase diagram.
Most promising tools for the experimental determination of these
characteristics are the energy scan programs at CERN, RHIC and at the upcoming
dedicated facilities of the third generation: FAIR and NICA.
The systematic analysis of higher moments of distributions of produced
particles in their dependence on the collision energy and the size of colliding 
systems shall provide answers to the above questions and allow for direct
comparison with predictions from the underlying theory, as provided by LQCD,
 see, e.g., Ref.~\cite{Karsch:2010ck}.

As long as the applicability of LQCD methods is bound to the region of
finite $T$ {and}  {$\mu_B/T \ll 1$}, any predictions for the
phase structure of QCD at high baryon densities including the possible
existence of critical points will rely on  effective {models}.
To  be relevant for the discussion of the above problems these {models} have
to share with QCD the property of chiral symmetry and its dynamical breaking
as well as a mechanism for confinement and deconfinement.

\begin{figure}[!thbp]
  \centering
    \includegraphics[width=8cm]{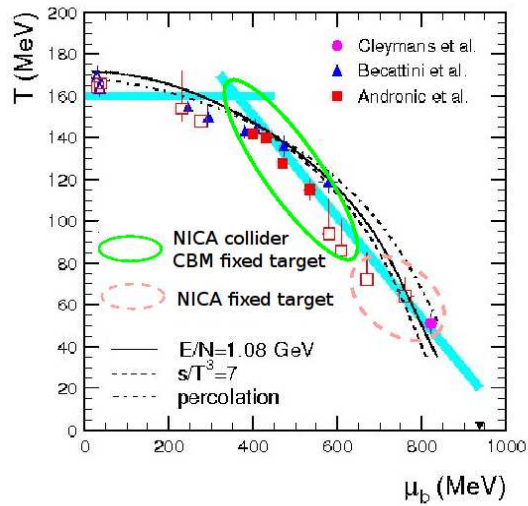}
  \caption{Freeze-out parameters from the statistical model
\cite{Andronic:2009gj} compared to phenomenological rules for explaining
their systematics in the temperature - chemical potential plane.
Ellipses indicate regions where dedicated future experiments (FAIR-CBM and
JINR-NICA) could provide new data for more detailed investigations.
}
    \label{fig:nica-freezeout}
\end{figure}

At present, the position of (pseudo-)critical lines in the  QCD phase diagram 
is constrained by  the  chemical freeze-out parameters ($T^{f},\mu_B^{f}$)
which have been obtained from the  statistical model analysis of  particle
yields obtained in heavy ion collisions
\cite{BraunMunzinger:2003zd,BraunMunzinger:2001ip}.
One of the most striking observations is the systematic behaviour of these
parameters with collision energy $\sqrt{s}$
\cite{Cleymans:1998fq,Cleymans:1999st,Cleymans:2005xv},
which has recently been given a simple parametric form \cite{Cleymans:2006qe}.
It has been observed, that the {resulting} freeze-out curve in the phase 
diagram is closely correlated to the thermodynamical quantities of the hadron
resonance gas described by the statistical model, see 
Fig.~\ref{fig:nica-freezeout}.
These phenomenological freeze-out {conditions}  make statements about the mean
energy per hadron $\langle E \rangle / \langle N \rangle \simeq 1.0$ GeV,
the dimensionless {entropy density $s/T^3\simeq 7$ and a total baryon and
antibaryon density $n_B + n_{\bar{B}} \simeq  0.12$ fm$^{-3}$.}
The freeze-out line provides a lower bound for the chiral restoration and
deconfinement transitions in the phase diagram.
Being coincident at low densities as inferred from LQCD \cite{Karsch:1994hm},
both transitions need not to occur simultaneously at high densities, thus 
allowing for an island of a quarkyonic phase \cite{McLerran:2007qj}
between hadron gas and quark-gluon plasma
with a (pseudo-)triple point \cite{Andronic:2009gj}.
The question appears for the physical mechanism which governs the chemical
freeze-out and which determines quantitatively the freeze-out parameters.
One aspect is provided by the requirement that hadrons should overlap in order
to facilitate flavor exchange reactions which establish chemical equilibrium.
This geometrical picture of freeze-out is successfully realized in a percolation
theory approach \cite{Magas:2003wi}.
Another aspect is the dynamical one: when the equation of state (EoS) possess
softest points (e.g., due to the dissociation of hadrons into their quark and
gluon constituents with a sufficient release of binding energy involved) which
is correlated with the freeze-out curve in the phase diagram, then it is
expected  that the hadron abundances are characterized by the corresponding $T$
and $\mu_B$ values \cite{Toneev:2000ym}.
The dynamical system got quasi trapped at the softest points for sufficient
time to achieve chemical equilibration before evaporating as a gas of hadron
resonances freely streaming to the particle detectors.
{Also the kinetic aspect of fast  chemical  equilibration was discussed in the
hadronic gas when accounting for a multi-hadron dynamics \cite{pbm}.}

All these mechanisms are appealing since they provide an intuitively clear 
picture, but they are flawed by the fact that their relation to fundamental 
aspects of the QCD phase transition like the chiral condensate as an order 
parameter are not an element of the description.

In the present contribution we develop an approach which relates the
geometrical as well as the hydrodynamic and kinetic aspects of chemical
freeze-out to the medium dependence of the chiral condensate.
We demonstrate within a beyond-meanfield extension
\cite{Schaefer:2007pw,Blaschke:2007np,Rossner:2007ik,Skokov:2010wb,Skokov:2010uh,Radzhabov:2010dd}
of the Polyakov NJL model
\cite{Fukushima:2003ib,Ratti:2005jh,Roessner:2006xn,Sasaki:2006ww}
how the excitation of hadronic resonances initiates the melting
of the chiral condensate which entails a Mott-Anderson type delocalization
\cite{MOTT:1968zz,Anderson:1978zz,Dobrosavljevic} of hadron wave functions
at the hadronic to quark matter phase transition \cite{Blaschke:1984yj}. 
Already for a schematic resonance gas consisting of pions and nucleons with
artificially enhanced numbers of degrees of freedom we can demonstrate that a
kinetic freeze-out condition for the above model provides quantitative
agreement with the phenomenological freeze-out curve.

We propose a generalization of this approach by including a more complete 
description of the chiral condensate which accounts for  contributions  from 
the hadron resonance gas and the chiral dynamics implemented from the PNJL 
model. We demonstrate, that in this approach  
correlations between freeze-out and chiral condensate get distorted.
We discuss, that this may be due to  our   approximations which exclude
the in-medium effects on hadron masses as well as the    hadron-hadron
interactions  which could result in 
delocalization of the hadron wave functions in accordance with the picture
of a Mott-Anderson transition.

\section{Mott-Anderson freeze-out and chiral condensate}

Recently, we have suggested, that the chemical freeze--out in heavy ion 
collisions can be linked to the chiral condensate \cite{Blaschke:2011ry}.
The starting point is based on the assumption, that  the chemical freeze--out 
sets in if the collision and the expansion time are equal. 
Thus, in the temperature-chemical potential $(T,\mu)$-plane, the  kinetic 
freeze-out condition  is  formulated as 
\begin{equation}
\tau_{\rm exp}(T,\mu)=\tau_{\rm coll}(T,\mu)~,
\label{fo}
\end{equation}
where $\tau_{\rm exp}(T,\mu)$ is the expansion time of the hadronic fireball, 
and 
\begin{equation}
\label{freezeout}
\tau_{\rm coll}^{-1}(T,\mu)=\sum_{i,j}\sigma_{ij}n_j~,
\end{equation}
is the inverse relaxation time for reactive collisions with $i,j=\pi, N, ...$ 
running over all particle species in the hadron resonance gas.
For the hadron-hadron cross sections $\sigma_{ij}$ we adopt the geometrical 
Povh-H\"ufner law,
\cite{Hufner:1992cu,Povh:1990ad}
\begin{equation}
\label{ph}
\sigma_{ij}=\lambda \langle r_i^2 \rangle \langle r_j^2 \rangle~,
\end{equation}
where $\lambda$ is a constant, being  of the order of the string tension,
$\lambda\sim 1~{\rm GeV/fm}= 5~{\rm fm}^{-2}$.
Note, that this behavior was also obtained for the quark exchange
contribution to hadron-hadron cross sections \cite{Martins:1994hd}.

A key point of our approach is that the radii of hadrons  depend on
$T$ and $\mu$  and that they  diverge, when hadron dissociation (Mott-Anderson
delocalization) driven basically by the restoration of the chiral symmetry,  
sets in. 
This has been quantitatively  studied for the pion \cite{Hippe:1995hu}, where
it has been shown, that close to the Mott-Anderson transition 
the pion radius is well approximated by
\begin{equation}
r_\pi^2(T,\mu)=\frac{3}{4\pi^2} F_\pi^{-2}(T,\mu)~.
\end{equation}
In addition, it has been demonstrated, that the GMOR relation holds out to the 
chiral phase transition, where pions merge the continuum of unbound quark matter
\cite{Blaschke:1999ab,Blaschke:2000gd}.
Furthermore, since the current-quark mass is  $T-$ and $\mu -$independent and 
the pion mass is ``chirally protected'',
the  $T-$ and  $\mu-$dependence of the chiral condensate should  be similar to 
that of the pion decay constant,
\begin{equation}
\label{GMOR}
F_\pi^2(T,\mu)=-m_0\langle \bar{q}q\rangle_{T,\mu}/m_\pi^2.
\end{equation}
Consequently, the relationship between the pion radius and the chiral 
condensate reads
\begin{equation}
r_\pi^2(T,\mu)=\frac{3m_\pi^2}{4\pi^2m_q}
|\langle \bar{q} q \rangle_{T,\mu}|^{-1}~.
\end{equation}
The Mott-Anderson delocalization of the pion wave function due to  melting
of the chiral condensate, as expressed in this formula, is the  important
element of the hadronic freeze-out  mechanism  suggested in
\cite{Blaschke:2011ry}.

For the nucleon, we  assume that its  radius consist of two
components; a medium independent hard core radius $r_0$, and a pion cloud
contribution as
\begin{equation}
r_N^2(T,\mu)=r_0^2+r_\pi^2(T,\mu)~,
\end{equation}
where from the vacuum values $r_\pi=0.59$ fm and $r_N=0.74$ fm  one gets
$r_0=0.45$ fm.

For the expansion time scale we adopt the  relation  which follows from
the entropy conservation, $S=s(T,\mu)~V(\tau_{\rm exp})={\rm const}$ with    
$V(\tau_{\rm exp})$ being a fireball volume. 
Assuming,  that
$V(\tau_{\rm exp})\propto \tau^3_{\rm exp}$ one gets
\begin{equation}
\tau_{\rm exp}(T,\mu)=a~s^{-1/3}(T,\mu)~,
\end{equation}
with $a$ being a constant of the order one.

To quantify the freeze-out conditions in the $(T-\mu)$ plane, we first consider 
a  medium which is composed of pions and nucleons only, however, with a 
variable number of degrees of freedom. In this  way, we account effectively 
for additional mesonic and baryonic stats in the hadron resonance gas.
A generalization to a medium composed of all hadrons with physical degrees of 
freedom will be discussed later.

We start from  a medium-modified chiral condensate obtained within the chiral 
perturbation theory \cite{Gasser:1986vb,Buballa:2003qv}, 
\begin{eqnarray}
\label{cond-new}
\frac{\langle\bar{q}q \rangle}{\langle\bar{q}q \rangle_{\rm vac}}&=&
1-\frac{n_{s,\pi}(T)}{m_\pi F_\pi^2}
-\frac{\Sigma_{\pi N} n_{s,N}(T,\mu)}{m_\pi^2F_\pi^2}~,
\end{eqnarray}
where  the pion-nucleon sigma-term is
$\Sigma_{\pi N}=m_0(\partial m_N/\partial m_0)=45$ MeV  \cite{Cohen:1991nk}, and
the scalar densities of pions $n_{s,\pi}$ and nucleons $n_{s,N}$ read
\begin{eqnarray}
n_{s,\pi}&=& d_\pi \int_0^\infty \frac{dp \, p^2}{2\pi^2} \frac{m_\pi}{E_\pi(p)}
f_\pi(E_\pi(p))~,\\
   n_{s,N}&=& \frac{d_N}{2\pi^2}\int_0^\infty dp \, p^2\,\frac{m_N}{E_N(p)}
\left\{f_N^+(E_N(p))+f_N^-(E_N(p)) \right\}~.
\end{eqnarray}
with $d_\pi$ the pionic and $d_N$ the nucleonic number of degrees of freedom. 
In the chiral limit and for $d_\pi=3$ we have, $n_{s,\pi}/m_\pi=T^2/8$.
\begin{figure}[!thbp]
  \centering
  \includegraphics[width=8cm]{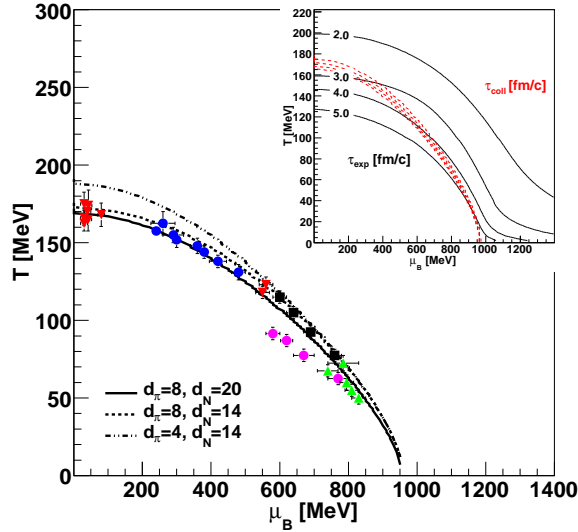}
  \caption{Freeze-out curves according to the kinetic freeze-out
condition, Eq.~(\ref{fo}), for the schematic resonance gas model with
different number of pion ($d_\pi$) and nucleon ($d_N$) degrees of freedom
compared to phenomenological values (symbols) \cite{Andronic:2009gj}.
The lines correspond to different choices for $d_\pi$ and $d_N$ given in the
legend. The inset shows the lines of constant expansion time (black solid) and
constant collision time (red dashed) for values 2, 3, 4, and 5 fm/c.
Lines for $\tau_{\rm coll}$ are so close together that their behavior
determines the location of hadronic freeze-out.}
    \label{fig:freezeout}
\end{figure}
Fig.~\ref{fig:freezeout}  shows  the chemical freeze-out lines in the 
$(T-\mu)$-plane obtained from the  kinetic freeze-out conditions (\ref{fo}) for 
three sets of pion and nucleon degrees of freedom.
The best description of the phenomenological freeze-out points is obtained for 
$d_\pi=8$ and  $d_N=20$.
The inset of Fig.~\ref{fig:freezeout}  shows a  dependence of the position of 
the freeze-out line on the expansion (solid-line) and the collision 
(dashed-line) times used in the model for $d_\pi=8$ and  $d_N=20$. 
 
From Fig.~\ref{fig:freezeout} one concludes, that the chemical freeze-out in 
the energy range between RHIC and GSI (SIS) occurs 3 - 5 fm/c after the
expansion of the collision fireball has started.
It is also evident from  Fig.~\ref{fig:freezeout}, that increasing $d_\pi$ 
results in decreasing   freeze-out temperatures in the meson-dominated region 
at small  $\mu_B/T$. On the other hand, increasing $d_N$ shifts the entire 
freeze-out curve towards lower $T$ and $\mu_B$.
The freeze-out curves calculated in this simplified model correspond to a 
reduction of the chiral condensate up to $50\%$ of its vacuum value, c.f. the 
upper right panel of Fig.~\ref{fig:cond_improved}. 
Thus, according to Eq.~(\ref{ph}),  the hadron-hadron cross section at 
freeze-out in (\ref{freezeout}) is roughly 3-4 times its  vacuum value.

In the next section, we introduce the  first results of extending the above 
freeze-out model including correctly all hadrons contained in the hadron 
resonance gas. We  start our discussion by introducing   a
generalization of the chiral condensate.

\section{Chiral condensate in a hadron resonance gas}

The chiral condensate is introduced based on the thermodynamical
potential  $\Omega(T,\mu)$  which is  decomposed into contributions from
 quarks and gluon (mean-field) and the hadronic (quantized) fluctuations in the
channels of Goldstone bosons ($G$ = $\pi$, $K$, $\eta$, $\eta'$), mesons
($M$ = $f_0$, $\rho$, $\omega$, ...) and baryons
($B$ = $N$, $\Lambda$, $\Sigma$, $\Delta$, ...),  
\begin{equation}
\label{omega}
\Omega(T,\mu)=\Omega_{\rm MF}(T,\mu)+\Omega_{\rm G}(T,\mu)
+\Omega_{\rm M}(T,\mu)+\Omega_{\rm B}(T,\mu)~.
\end{equation}

For the mean-field contribution in the quark and gluon sector we employ the 
expression obtained within the  PNJL model\footnote{We use the
approximation $\bar\Phi = \Phi = [ 1 + 2\,\cos(\bar\phi_3/T) ]/3$ for the
mean value of the traced Polyakov loop.} \cite{Hansen:2006ee},
\begin{eqnarray}
\Omega_{\rm MF}(T,\mu) &=&
- 2\sum_{f}\left\{\int \frac{d^3p}{(2\pi)^3} \left\{3 \varepsilon_f
+  T\; \ln \left[N_\Phi^{+}(\varepsilon_f)
N_{\bar{\Phi}}^{-}(\varepsilon_f) \right]\right\}
+ \frac{\sigma_f^2}{8G} - \frac{\omega_f^2}{8G_V}\right\}
+ {\cal{U}}(\Phi),
\label{ommf}
\end{eqnarray}
where
$N_\Phi^{\pm}(\varepsilon)=1 + 3 \Phi e^{-\beta (\varepsilon\mp\mu)}
+ 3\Phi e^{-2\beta (\varepsilon \mp\mu)} + e^{-3\beta (\varepsilon\mp\mu)}$,
with $\varepsilon_f(p)=\sqrt{p^2+m_f^2(T,\mu)}$ being the quasi-particle
energy  and $m_f(T,\mu)=m_{0,f}+\sigma_f(T,\mu)$  the dynamically
generated mass of quark with  flavor $f$.
The vector fields $\omega_f$ renormalize the quark chemical potentials
$\mu_f=\mu_{f,0}-\omega_f$.
For the Polyakov-loop potential we take the logarithmic form motivated by the
SU(3) Haar measure,
\begin{equation}
{\cal{U}}(\Phi ,T) = \left[-\,\frac{1}{2}\, a(T)\,\Phi^2 \;+\;b(T)\, \ln(1 -
6\, \Phi^2 + 8\, \Phi^3 - 3\, \Phi^4)\right] T^4 \ ,
\end{equation}
with the corresponding parameters $a(T)$ and $b(T)$ from 
Ref.~\cite{Rossner:2007ik}.
A possible dependence of the $T_0$  on the number of active flavors and on the 
chemical potential has been suggested in  Ref.~\cite{Schaefer:2007pw}.
Here we  use the constant value of $T_0=187$ MeV appropriate for the  
(2+1)--flavor model.

In the following we consider only the on-mass-shell hadron contributions
and neglect the continuum correlations.
This approximation is motivated by the fact that
hadronic freeze-out occurs in a region of the phase diagram which
does not exceed the limits of the hadronic phase.
It may, however, turn out to be insufficient in the vicinity of the chiral 
restoration transition where the energy gap between hadronic bound states
and the threshold for the quark continuum states becomes too small to suppress
the excitation of the latter relative to the former.

The Goldstone and meson contributions to the thermodynamical potential are
given as
\begin{eqnarray}
\label{meson-omega}
\Omega_{G,M}(T,\mu)=\sum_{\stackrel{G=\pi,...}{M=f_0,...}} d_{G,M}
\int \frac{d^3k}{(2\pi)^3} \left\{\frac{E_{G,M}(k)}{2}
+  T\; \ln \left[1 - e^{-\beta E_{G,M}(k)}\right]\right\}~,
\end{eqnarray}
where the index $G$ ($M$) denotes the actual meson with degeneracy factor
$d_G$ ($d_M$) and the dispersion $E_{G,M}(k)=\sqrt{k^2+M_{G,M}^2}$.

The contribution of baryons to the partition function is considered as a
quasiparticle Fermi gas
\begin{eqnarray}
\label{baryon-omega}
\Omega_{B}(T,\mu)= -\sum_{B=N,...} d_B
\int \frac{d^3k}{(2\pi)^3} \left\{E_B(k)
+  T\; \ln \left\{\left[1 + e^{-\beta (E_B(k)-\mu_B)}\right]
\left[1 + e^{-\beta (E_B(k)+\mu_B)}\right]\right\}\right\}.
\end{eqnarray}

In this work we  study  flavor-symmetric matter where quantities for the sector 
of the light flavors $u$ and $d$ are  denoted by the subscript $q$. 
We define, $\sigma_q=(\sigma_u+\sigma_d)/2$, while
$m=m_u=m_d$ and  $m_0=m_{0,u}=m_{0,d}$.
The light quark condensate is defined as
\begin{eqnarray}
\label{cond}
\langle\bar{q}q \rangle&=&\langle\bar{u}u + \bar{d}d \rangle
=\left(\frac{\partial}{\partial m_{0,u}}+\frac{\partial}{\partial m_{0,d}}\right)
\Omega(T,\mu)~.
\end{eqnarray}
In this work we will not yet solve selfconsistently the gap equations for the
set of order parameters $\{\sigma_f\}, \{\omega_f\}, ...$ which correspond to
the thermodynamical equilibrium state characterized by the global minimum of
the thermodynamical potential (\ref{omega}), i.e. including contributions from
hadronic resonances.
Such contributions  have  been discussed for mesonic fluctuations in
\cite{Hufner:1994ma,Zhuang:1994dw} for the NJL model and in
\cite{Hansen:2006ee,Hell:2008cc} for the PNJL model.
Also recently mesonic fluctuations were included in these models within the
functional renormalization group approach \cite{rg1,rg2}.

To get  $\partial m_{M,B}/\partial m_{0,f}$ we  use rather
simplified mass formulas
\begin{equation}
\label{masses}
m_B=(3-N_s)(\sigma_q + m_{0})+N_s(\sigma_s+m_{0,s})+\kappa_B~~,~~
m_M=(2-N_s)(\sigma_q + m_{0})+N_s(\sigma_s+m_{0,s})+\kappa_M~,
\end{equation}
where $\kappa_B$ and $\kappa_M$ are state-dependent constants for baryons and
mesons, respectively.

Let us first discuss  a derivation of the two-quark condensate from 
Eq.~(\ref{cond}) 
including contributions, beyond the quark mean-field, of  mesons and baryons
forming the hadron resonance gas.
To this end, we employ the schematic mass formulas for baryons and mesons
(\ref{masses}), the GMOR motivated relationship
 \begin{eqnarray}
\label{GMOR-G}
\left(\frac{\partial}{\partial m_{0,u}}+\frac{\partial}{\partial m_{0,d}}\right)
m_G^2=- r_G \langle\bar{q}q \rangle / f_G^2
\end{eqnarray}
and assume flavor symmetry of the vacuum
$\langle\bar{s}s \rangle \approx \langle\bar{u}u + \bar{d}d \rangle/2$.
For the coefficients $r_G$, see Table \ref{tab:Gold} and
Ref.~\cite{Leupold:2006ih}.
\begin{table}
\centering
\begin{tabular}{c|c|c|c}
$G$&$F_G$ [MeV]&$d_G$&$r_G$\\
\hline
$\pi$&92.4&3&1\\
$K$&113&4&1/2\\
$\eta$&124&1&1/3\\
$\eta'$&107&1&2/3\\
\hline
\end{tabular}
\caption{Parameter values for Goldstone bosons, see also \cite{Leupold:2006ih}.
\label{tab:Gold}}
\end{table}
We obtain
\begin{eqnarray}
\label{cond2}
\langle \bar{q}q\rangle&=&
-4 N_c
\int \frac{dp \, p^2}{2\pi^2} \frac{m}{\varepsilon_p}
\left[1- f_\Phi^+ -f_\Phi^-\right]
- \sum_{G=\pi,K,\eta,\eta'} d_G r_G \int \frac{dp \, p^2}{2\pi^2}
\frac{\langle \bar{q}q\rangle}{2c_G F_\pi^2E_G(p)}
\left[\frac{1}{2}+f_G(E_G(p)) \right]\nonumber\\
&&+ \sum_{M=f_0,\omega, ...} d_M (2-N_s^M)
\int \frac{dp \, p^2}{2\pi^2} \frac{m_M}{E_M(p)}
\left[\frac{1}{2}+f_M(E_M(p)) \right]\nonumber\\
&&- \sum_{B=N,\Lambda, \dots} d_B (3-N_s^B)
\int \frac{dp \, p^2}{2\pi^2} \frac{m_B}{E_B(p)}
\left[1-f_B^+(E_B(p)) -f_B^-(E_B(p))\right]~,
\end{eqnarray}
where $c_G=F_G^2/F_\pi^2$ and 
the PNJL model  distribution functions  for   quarks read
\begin{equation}
\label{f-PNJL}
f^\pm_\Phi(\varepsilon)=\{\Phi[ e^{-\beta (\varepsilon\mp \mu)}
+ 2 e^{-2\beta (\varepsilon \mp \mu)}] + e^{-3\beta (\varepsilon \mp\mu)}\}
/N_\Phi^\pm(\varepsilon)~.
\end{equation}
The values of the Polyakov loop $\Phi$ in the phase diagram are found from a
similar gap equation corresponding to the solution of the extremum condition
$\partial \Omega/\partial \Phi=0$.
The distribution functions for mesons and Goldstone bosons as bound states of
quark with flavor $f_1$ and antiquark with flavor $f_2$ are:
$f_{M,G}=1/\{\exp[\beta(E_{M,G}-(\mu_{f_1}-\mu_{f_2}))]-1\}$.
The distribution function for baryons composed of three quarks with flavors
$f_1$, $f_2$ and $f_3$ are:
$f_{B}^\pm=1/\{\exp[\beta(E_{B}\mp\mu_{f_1}\mp\mu_{f_2}\mp\mu_{f_3})]+1\}$.

We will use now the property of chiral protection of the pion mass in the
phase with broken chiral symmetry which entails that the in-medium modification
of the chiral condensate shall be compensated by that of the squared pion
decay constant
\begin{equation}
\frac{\langle \bar{q}q\rangle}{F_\pi^2}=
\frac{\langle \bar{q}q\rangle_{\rm vac}}{F_{\pi,{\rm vac}}^2}
=-\frac{m_\pi^2}{m_0}~.
\end{equation}
In the following we  drop the subscript index ``$\rm vac$'' and use $F_\pi$ for
the vacuum value of the  pion decay constant.
The final form for the in-medium modification of the chiral condensate is then
\begin{eqnarray}
\label{condfin}
\frac{\langle \bar{q}q\rangle}{\langle \bar{q}q\rangle_{\rm vac}}&=&
1-\frac{m_0}{F_\pi^2m_\pi^2}
\bigg\{4 N_c \int \frac{dp \, p^2}{2\pi^2} \frac{m}{\varepsilon_p}
\left[f_\Phi^+ +f_\Phi^-\right]
\nonumber\\
&& +\sum_{M=f_0,\omega, ...} d_M (2-N_s)\int\frac{dp\, p^2}{2\pi^2}
\frac{m_M}{E_M(p)}f_M(E_M(p))
\nonumber\\
&&+ \sum_{B=N,\Lambda, \dots} d_B (3-N_s)
\int \frac{dp \, p^2}{2\pi^2} \frac{m_B}{E_B(p)}
\left[f_B^+(E_B(p))+f_B^-(E_B(p))\right]\bigg\}
\nonumber\\
&&-\sum_{G=\pi,K,\eta,\eta'} \frac{d_G r_G}{4\pi^2 F_G^2}
\int dp \, \frac{p^2 }{E_G(p)}f_G(E_G(p))~.
\end{eqnarray}
The above  result for the two-quark condensate  reproduces that found in 
Ref.~\cite{Leupold:2006ih} for the hadronic resonance contributions, and
generalizes it by including the quark mean-field contribution according to the
chiral quark dynamics from the PNJL model.

\section{Results of the PNJL-hadron resonance gas model}
In the following,   we summarize first results for the chiral condensate,
which is  the main input  to our description of the freeze-out conditions 
within the Mott-Anderson hadron localization mechanism.
For these  exploratory studies,  we neglect the hadron effective mass and the
renormalization of the chemical potentials, so that we stay rather close to
the standard statistical model approach which operates without medium
modification of the hadronic states.

We use standard values of the NJL model parametrization, e.g., from
\cite{Grigorian:2006qe}, with $G\Lambda^2=2.316$, $\Lambda=602.3$ MeV,
$m_0=5.5$ MeV, $m_{s,0}=138.7$ MeV, $M_\pi=140$ MeV, $M_K=495$ MeV,
$f_\pi=92.4$ MeV and $f_K=93.6$ MeV.

\begin{figure}[hbtp]
  \centering
    \includegraphics[width=13cm]{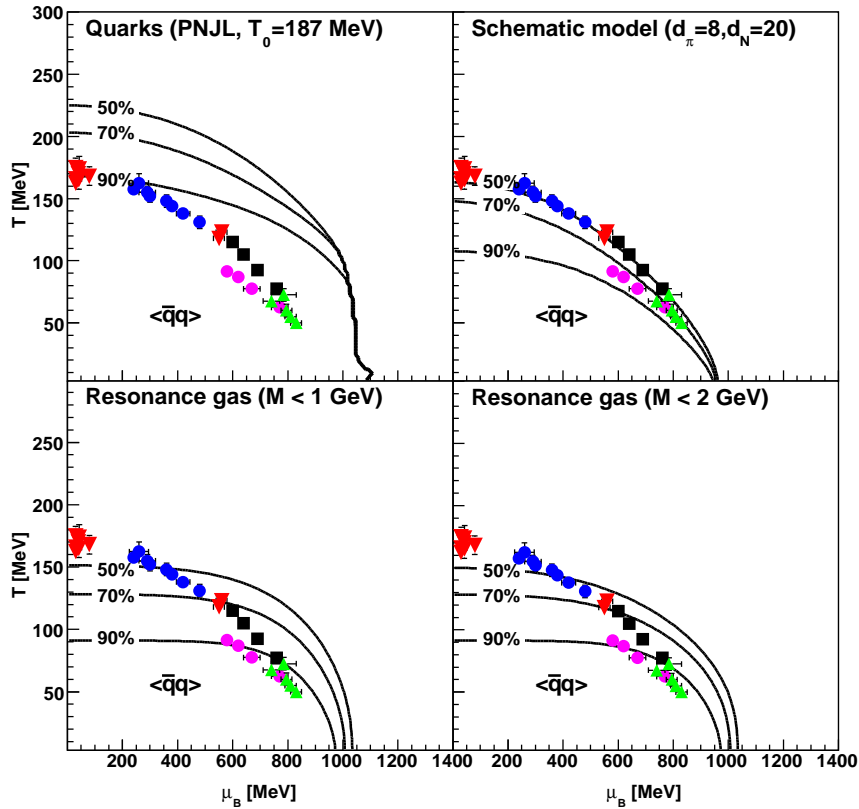}
  \caption{Chiral condensate in the $T$-$\mu$ plane compared to freeze-out
data. Upper left: two-quark condensate for PNJL quark meanfield without
hadronic contributions; upper right: schematic resonance gas model
\cite{Blaschke:2011ry}; lower left: present work (\ref{condfin})
excluding quark meanfield \cite{Leupold:2006ih} including hadronic states with
masses up to 1 GeV; lower right: the same for masses up to 2 GeV.
\label{fig:cond_improved}
}
\end{figure}

In the first step, we confirm  consistency with our model assumption,
that quark degrees of freedom do not yet interfere at the freeze-out
temperatures and chemical potentials.
In order to do that we neglect in the in-medium condensate
(\ref{condfin})  contributions from Goldstones as well as  mesons
and baryons by setting the corresponding degrees of freedom to
zero, $d_M=d_B=d_G=0$.
The remaining expression is that for the PNJL model condensate in the 
mean-field approximation,
\begin{eqnarray}
\frac{\langle \bar{q}q\rangle}{\langle \bar{q}q\rangle_{\rm vac}}&=&
1-\frac{m_0 n_{s,q}}{F_\pi^2m_\pi^2}~~,~~
n_{s,q}= 4 N_c \int \frac{dp \, p^2}{2\pi^2} \frac{m}{\varepsilon_p}
\left[f_\Phi^+ +f_\Phi^-\right]~.
\end{eqnarray}
The results for the freeze-out lines  with the above condensate  are  
shown in the upper-left plot of Fig.~\ref{fig:cond_improved}. They 
confirm,  that the reduction of the vacuum condensate is still by less than
10\% along the chemical freeze-out data.
Since the modification of the mean-field contribution stemming from the quark
excitations is proportional to $f_\Phi$, it is effectively suppressed by
the Polyakov-loop, compared to the standard NJL model case.
%

We  investigate further the role of the hadronic states for the behavior of the
chiral condensate.
First we consider a schematic model  by reducing a complete  set of states to
nucleons and pions only. Their numbers of degrees of freedom, $d_\pi=8$ and
$d_N=14$ are  chosen such that to get the best description of the  freeze-out 
data. We observe that the freeze-out curve is correlated with the condensate
reduction to 50 \%.


The lower panels of Fig.~\ref{fig:cond_improved}  show  results for
the quark condensate from  Eq.~(\ref{condfin})  when the
contribution of the quark mean-field is excluded. 
This assumption   corresponds to
the two-quark condensate form  \cite{Leupold:2006ih}, including hadronic states
with masses up to 1 GeV (lower left) and up to 2 GeV (lower right).
We observe that the shape of the lines of constant quark condensate in the
($T$- $\mu$)- plane is different from the systematics of the freeze-out curve
and thus different from the behavior of the schematic resonance gas model.
Extending the mass limit of hadronic states from 1 GeV to 2 GeV has only a
minor influence on the onset of the chiral restoration (reduction up to 50\%),
visible in the region of  large  baryon-chemical potentials, since in this
mass range predominantly baryonic states are added.
Obviously, the straightforward generalization of the simple kinetic freeze-out
model based on a scaling of hadron cross sections with the chiral condensate
towards a realistic resonance gas model requires improvements of the
assumptions made in the present work.
We will discuss them in the concluding section.

\section{Conclusion and outlook}

We have developed a chemical freeze-out mechanism which accounts  for a strong
medium dependence of the rates for inelastic flavor-equilibrating collisions. 
We based our concept on the delocalization of hadronic wave functions and 
growing hadronic radii when approaching the chiral restoration.
This approach relates the geometrical (percolation) as well as the
hydrodynamic (softest point) and kinetic (quark exchange) aspects of chemical
freeze-out to the medium dependence of the chiral condensate.
For our model calculations we have employed a beyond-mean-field, {\it effective}
extension of the Polyakov-loop NJL model.
We could demonstrate how the excitation of hadronic resonances initiates the
melting of the chiral condensate which entails a Mott-Anderson type
delocalization of the hadron wave-functions and a sudden drop in the relaxation
time for flavor equilibration.

For a schematic resonance gas consisting of pions and nucleons with
variable numbers of degrees of freedom we could demonstrate, that
the kinetic freeze-out condition {can provide} quantitative agreement with the
phenomenological freeze-out curve for suitably chosen values of the degeneracy
of pionic and nucleonic states.

We have developed this approach further by replacing the schematic resonance
gas model for the chiral condensate with a full treatment of hadronic
resonances and studied  their influence on the chiral condensate.
To this end generalized models for hadron resonance masses were required,
which reveal their dependence on the current quark mass like, e.g., the model
of Leupold \cite{Leupold:2006ih} which we have adopted here.

The obtained result for  the chiral condensate which triggers the
freeze-out in the Mott-Anderson localization model, showed that the
approach by Leupold to the chiral condensate in a hadron resonance gas, 
does not show the expected correlation between chiral condensate and freeze-out.

We see the main reason for the failure of the improved model in  the
inconsistency of the Mott-Anderson delocalization picture with the assumption
of on-shell hadronic spectral functions in the  thermodynamical
potential (\ref{omega}) as well as in the very crude baryon and meson mass 
formulae (\ref{masses}).
The latter provides a value for the pion-nucleon-sigma term which is about a
factor three too low. Clearly, the effects of chiral hadron-hadron interactions
are missing in (\ref{omega}).
At high phase-space occupation, one expects, that due to the finite
hadronic radii the overlap of their wave functions should lead to effects of
the Pauli principle on the quark level (cf. excluded volume) and to the 
appearance of multi-quark admixtures to the hadronic states 
\cite{Pirner:1980eu,Pirner:2010fw}.
When interpreted in terms of hadron-hadron scattering phase shifts
(see, e.g.,
Refs.~\cite{Barnes:1991em,Blaschke:1992qa,Barnes:1993nu,Martins:1994hd} for
early examples of quark exchange contributions)
the quark substructure effects could be accounted for by a Beth-Uhlenbeck type
generalization of the thermodynamical potential
\cite{Beth:1937,Schmidt:1990,Hufner:1994ma}.
It is expected, that besides changes in the hadron wave functions, the masses
of baryons shall get reduced at high densities whereas the masses of mesons
could increase toward the chiral restoration. Such a behavior would distort the
lines of constant chiral condensate towards the freeze-out data.
Here, we have disregarded the in-medium modifications of hadronic masses, 
which are currently being under consideration.

\begin{acknowledgements}
D.B. and K.R. acknowledge support from the Polish Ministry of Science and
higher Education (MNiSW) under grant Nos. NN 202 23 1837. 
The work of D.B. was also supported by CompStar, a Research Networking
Programme of the European Science Foundation, by the MNiSW grant CompStar-POL
and by the Russian Fund for Basic Research under grant No. 11-02-01538-a.
\end{acknowledgements}


\begin{thebibliography}{3}
%
%
\bibitem{Karsch:2010ck}
  F.~Karsch and K.~Redlich,
  Phys.\ Lett.\  B {\bf 695}, 136 (2011).

\bibitem{BraunMunzinger:2003zd}
  P.~Braun-Munzinger, K.~Redlich and J.~Stachel,
  in: {\it Quark Gluon Plasma 3}, Eds. R.C. Hwa and Xin-Nian Wang,
  World Scientific (2003), pp. 491-599;  [arXiv:nucl-th/0304013].

\bibitem{BraunMunzinger:2001ip}
  P.~Braun-Munzinger, D.~Magestro, K.~Redlich and J.~Stachel,
  Phys.\ Lett.\  B {\bf 518}, 41 (2001).

\bibitem{Cleymans:1998fq}
  J.~Cleymans, K.~Redlich,
  Phys.\ Rev.\ Lett.\  {\bf 81}, 5284-5286 (1998).

\bibitem{Cleymans:1999st}
  J.~Cleymans, K.~Redlich,
  Phys.\ Rev.\  {\bf C60}, 054908 (1999).

\bibitem{Cleymans:2005xv}
  J.~Cleymans, H.~Oeschler, K.~Redlich and S.~Wheaton,
  Phys.\ Rev.\  C {\bf 73}, 034905 (2006).

\bibitem{Cleymans:2006qe}
  J.~Cleymans, H.~Oeschler, K.~Redlich and S.~Wheaton,
  J.\ Phys.\ G {\bf 32}, S165 (2006).

\bibitem{Karsch:1994hm}
  F.~Karsch and E.~Laermann,
  Phys.\ Rev.\  D {\bf 50}, 6954 (1994).

\bibitem{McLerran:2007qj}
  L.~McLerran and R.~D.~Pisarski,
  Nucl.\ Phys.\  A {\bf 796}, 83 (2007).

\bibitem{Andronic:2009gj}
  A.~Andronic, D.~Blaschke, P.~Braun-Munzinger {\it et al.},
  Nucl.\ Phys.\  {\bf A837}, 65-86 (2010).

\bibitem{Magas:2003wi}
  V.~Magas and H.~Satz,
  Eur.\ Phys.\ J.\  C {\bf 32}, 115 (2003).

\bibitem{Toneev:2000ym}
  V.~D.~Toneev, J.~Cleymans, E.~G.~Nikonov {\it et al.},
  J.\ Phys.\ G {\bf G27}, 827-832 (2001).

\bibitem{pbm}
 P.~Braun-Munzinger, J.~Stachel and C.~Wetterich,
  Phys.\ Lett.\  B {\bf 596}, 61 (2004)

\bibitem{Schaefer:2007pw}
  B.~J.~Schaefer, J.~M.~Pawlowski and J.~Wambach,
  Phys.\ Rev.\  D {\bf 76}, 074023 (2007).

\bibitem{Blaschke:2007np}
  D.~Blaschke, M.~Buballa, A.~E.~Radzhabov and M.~K.~Volkov,
  Yad.\ Fiz.\  {\bf 71}, 2012 (2008).

\bibitem{Rossner:2007ik}
  S.~Roessner, T.~Hell, C.~Ratti and W.~Weise,
  Nucl.\ Phys.\  A {\bf 814}, 118 (2008).


\bibitem{Skokov:2010wb}
  V.~Skokov, B.~Stokic, B.~Friman and K.~Redlich,
  Phys.\ Rev.\  C {\bf 82}, 015206 (2010).

\bibitem{Skokov:2010uh}
  V.~Skokov, B.~Friman, K.~Redlich,
  Phys.\ Rev.\  {\bf C83}, 054904 (2011).
  [arXiv:1008.4570 [hep-ph]].

\bibitem{Radzhabov:2010dd}
  A.~E.~Radzhabov, D.~Blaschke, M.~Buballa and M.~K.~Volkov,
  Phys.\ Rev.\  {\bf D83}, 116004 (2011).

\bibitem{Fukushima:2003ib}
  K.~Fukushima,
  Prog.\ Theor.\ Phys.\ Suppl.\  {\bf 151} (2003) 171;
%
  Phys.\ Lett.\  B {\bf 591}, 277 (2004);
%
  Phys.\ Rev.\  D {\bf 68}, 045004 (2003);
%
  Phys.\ Lett.\  B {\bf 553}, 38 (2003).

\bibitem{Ratti:2005jh}
  C.~Ratti, M.~A.~Thaler and W.~Weise,
  Phys.\ Rev.\  D {\bf 73} (2006) 014019.

\bibitem{Roessner:2006xn}
  S.~Roessner, C.~Ratti and W.~Weise,
  Phys.\ Rev.\  D {\bf 75}, 034007 (2007).

\bibitem{Sasaki:2006ww}
  C.~Sasaki, B.~Friman and K.~Redlich,
  Phys.\ Rev.\  D {\bf 75}, 074013 (2007).

\bibitem{MOTT:1968zz}
  N.~F.~Mott,
  Rev.\ Mod.\ Phys.\  {\bf 40}, 677 (1968).

\bibitem{Anderson:1978zz}
  P.~W.~Anderson,
  Rev.\ Mod.\ Phys.\  {\bf 50}, 191-201 (1978).

\bibitem{Dobrosavljevic}
  V.~Dobrosavljevi\'c,
  Int. J. Mod. Phys. B {\bf 24}, 1680 (2010).

\bibitem{Blaschke:1984yj}
  D.~Blaschke, F.~Reinholz, G.~R\"opke and D.~Kremp,
  Phys.\ Lett.\  B {\bf 151}, 439 (1985).

\bibitem{Blaschke:2011ry}
  D.~B.~Blaschke, J.~Berdermann, J.~Cleymans, K.~Redlich,
  Heavy Ions {\bf 1}, in press (2011);
  [arXiv:1102.2908 [nucl-th]].

\bibitem{Hufner:1992cu}
  J.~H\"ufner and B.~Povh,
  Phys.\ Rev.\  D {\bf 46}, 990 (1992).

\bibitem{Povh:1990ad}
  B.~Povh, J.~H\"ufner,
  Phys.\ Lett.\  B {\bf 245}, 653 (1990).

\bibitem{Martins:1994hd}
  K.~Martins, D.~Blaschke and E.~Quack,
  Phys.\ Rev.\  C {\bf 51}, 2723 (1995).


\bibitem{Hippe:1995hu}
  H.~J.~Hippe and S.~P.~Klevansky,
  Phys.\ Rev.\  C {\bf 52}, 2172 (1995).

\bibitem{Blaschke:1999ab}
  D.~Blaschke and P.~C.~Tandy,
  in: {\it Understanding Deconfinement in QCD}, World Scientific (2000),
  pp. 218-230; [arXiv:nucl-th/9905067].

\bibitem{Blaschke:2000gd}
  D.~Blaschke, G.~Burau, Yu.~L.~Kalinovsky, P.~Maris and P.~C.~Tandy,
  Int.\ J.\ Mod.\ Phys.\  A {\bf 16}, 2267 (2001).

\bibitem{Gasser:1986vb}
  J.~Gasser and H.~Leutwyler,
  Phys.\ Lett.\  B {\bf 184}, 83 (1987).

\bibitem{Buballa:2003qv}
  M.~Buballa,
  Phys.\ Rept.\  {\bf 407}, 205 (2005).

\bibitem{Cohen:1991nk}
  T.~D.~Cohen, R.~J.~Furnstahl and D.~K.~Griegel,
  Phys.\ Rev.\  C {\bf 45}, 1881 (1992).


\bibitem{Hansen:2006ee}
  H.~Hansen, W.~M.~Alberico, A.~Beraudo, A.~Molinari, M.~Nardi and C.~Ratti,
  Phys.\ Rev.\  D {\bf 75}, 065004 (2007).


\bibitem{Hufner:1994ma}
  J.~H\"ufner, S.~P.~Klevansky, P.~Zhuang and H.~Voss,
  Annals Phys.\  {\bf 234}, 225 (1994).

\bibitem{Zhuang:1994dw}
  P.~Zhuang, J.~H\"ufner and S.~P.~Klevansky,
  Nucl.\ Phys.\  A {\bf 576}, 525 (1994).

\bibitem{Hell:2008cc}
  T.~Hell, S.~Roessner, M.~Cristoforetti and W.~Weise,
  Phys.\ Rev.\  D {\bf 79}, 014022 (2009).

\bibitem{rg1}
See, e.g., J.~Berges, N.~Tetradis and C.~Wetterich, 
Phys.\ Rept.\  {\bf 363}, 223 (2002).

\bibitem{rg2}
B.~J.~Schaefer, J.~M.~Pawlowski and J.~Wambach,
  Phys.\ Rev.\ D {\bf 76}, 074023 (2007);\\
 B.~Stokic, B.~Friman and K.~Redlich,
  Eur. Phys. J. C {\bf 67},  425 (2010);\\
  E.~Nakano, B.~J.~Schaefer, B.~Stokic, B.~Friman and K.~Redlich,
  Phys.\ Lett.\  B {\bf 682}, 401 (2010);\\
  V.~Skokov, B.~Friman and K.~Redlich,
  Phys.\ Rev.\  {\bf C83}, 054904 (2011);\\
  T.~K.~Herbst, J.~M.~Pawlowski, B.~-J.~Schaefer,
  Phys.\ Lett.\ B {\bf 696}, 58 (2011).

\bibitem{Leupold:2006ih}
     S.~Leupold,
     J. \ Phys.\ G {\bf 32}, 2199 (2006).

\bibitem{Grigorian:2006qe}
  H.~Grigorian,
  Phys.\ Part.\ Nucl.\ Lett.\  {\bf 4}, 223 (2007).

\bibitem{Pirner:1980eu}
  H.~J.~Pirner, J.~P.~Vary,
  Phys.\ Rev.\ Lett.\  {\bf 46}, 1376 (1981).

\bibitem{Pirner:2010fw}
  H.~J.~Pirner, J.~P.~Vary,
  Phys.\ Rev.\  {\bf C84}, 015201 (2011).

\bibitem{Barnes:1991em}
  T.~Barnes, E.~S.~Swanson,
  Phys.\ Rev.\  {\bf D46}, 131-159 (1992).

\bibitem{Blaschke:1992qa}
  D.~Blaschke, G.~R\"opke,
  Phys.\ Lett.\  {\bf B299}, 332-337 (1993).

\bibitem{Barnes:1993nu}
  T.~Barnes, S.~Capstick, M.~D.~Kovarik, E.~S.~Swanson,
  Phys.\ Rev.\  {\bf C48}, 539-552 (1993).

\bibitem{Beth:1937}
 E.~Beth, G.~E.~Uhlenbeck,
  Physica {\bf 4}, 915 (1937).

\bibitem{Schmidt:1990}
  M.~Schmidt, G.~R\"opke, H.~Schulz,
  Ann. Phys. {\bf 202}, 57 (1990).

\end{thebibliography}


\end{document}